\documentstyle[eqsecnum,aps]{revtex}

\newcommand{\bml}{\begin{mathletters}}
\newcommand{\eml}{\end{mathletters}}

\begin{document}

\draft

\title{Canonical Quantization for the Light-Front Weyl Gauge}
 
\author{
{\it Jerzy A. Przeszowski} \\
Institute of Fundamental Technological Research \\ 	
Polish Academy of Science\\
\'Swi\c{e}tokrzyska 21, 00-049 Warsaw, Poland \\ }

\date{\today}

\maketitle

\begin{abstract}
The canonical quantization on a single light front is performed for
the Abelian gauge fields with the Weyl gauge $A_{+} = A^{-} = 0$
coupled with fermion field currents.  The analysis is carried
separately for 1+1 dimensions and for higher dimensions. 
The Gauss law, implemented weakly as the condition on states,
selects physical subspace with the Poincar\'{e} covariance
recovered. The perturbative gauge field propagators are found
with the ML prescription for their spurious poles. The LF
Feynman rules are found and their equivalence with the usual
equal-time perturbation for the S-matrix elements is studied
for all orders. 
\end{abstract}

\pacs{11.15.Bt, 11.10.K.k, 12.20.Ds}
\

\section{Introduction} 

In the last decade, there is a resurgence of interest in the
formulation of quantum field theory on a null-plane surface of
$x^{+} = const$, called after Dirac \cite{Dirac}, a light-front
(LF) \cite{BrodPaulPin1998}. 
Though nowadays the basic motivation for it lies in the
non-perturbative Hamiltonian field theory \cite{Wilson}, there
are still some problems with a consistent formulation of the
perturbative LF calculations. In this paper the perturbative
approach to the quantum electrodynamics will be discussed with a
special attention paid to the Abelian gauge vector fields.\\
Usually, the LF quantization is performed with the light-cone
(LC) gauge condition $A_{-} = A^{+} = 0$ being imposed on gauge
fields, this removes all non-physical gauge modes from the
dynamical system and introduces instantaneous (in $x^{+}$)
interactions of currents in the Hamiltonian \cite{Kogut1968}.
All these effectively lead to the perturbative gauge field
propagator with the Cauchy Principal Value (CPV) prescription
for the {\em spurious pole} at $k_{-} = 0$
\begin{equation}
D^{LC}_{\mu \nu}(x) = i \left < 0 \left | T A_{\mu } (x) A_\nu(0)
\right| 0 \right > =  \int \frac{d^4k}{(2\pi)^4} \frac{e^{- i k
\cdot x}}{2 k_{+} k_{-} - k_\perp^{2} + i \epsilon} \left[g_{\mu
\nu} - \frac{n^{LC}_\mu k_\nu + n^{LC}_\nu
k_\mu}{\left[k_{-}\right]_{CPV}}\right],
\label{CPVAAprop} 
\end{equation}
where the LC gauge vector $n^{LC}_\mu= (n^{LC}_{+} = 1, n^{LC}_{-}
= n^{LC}_i = 0)$ chooses the LC gauge $n^{LC}\cdot A = A_{-} = 0$.
Unfortunately this result is inconsistent with the equal-time
canonical quantization \cite{Bassetto1985}, 
where the LC gauge allows for the propagation of non-physical
modes and leads to the gauge field propagator with the causal
non-covariant distribution $\left[k_{-}\right]_{ML} = (k_{-} +
i \epsilon {\rm sgn}{k_{+}})^{-1}$, so-called the
Mandelstam-Leibbrandt (ML) prescription \cite{ML}. Because 
the perturbative calculations of the gauge-invariant Wilson loops 
\cite{Physical and Nonstandard Gauges} uniquely indicates 
the ML prescription as the only consistent regularization, then
one finds that the LF canonical quantization is perturbatively
inconsistent for the LC gauge.\\
Recent LF canonical attempts, where one uses two light-fronts
$x^{+} = const.$ and $x^{-} = const.'$ as the quantization 
surfaces \cite{McCartorRob}, \cite{Soldati} were only partially
succesful. For the free field model, the ML prescription has
appeared in the chronological product of gauge fields, but one
can doubt if this approach can be consistently implemented for
interacting fields. We believe that another, more fundamental 
solution to this problem has to be found. \\
One possibility is to choose another null axial gauge condition
$A_{+}= A^{-} = 0$, which for the reasons that we will explain
later, we call the light-front Weyl (LF-Weyl) gauge. In the
equal-time formulation both null axial gauges are treated on
equal footing, therefore we expect the free gauge field
propagator in the form
\begin{equation}
D^{Weyl}_{\mu \nu}(x) = i \left < 0 \left | T A_{\mu } (x) A_\nu(0)
\right| 0 \right > =  \int \frac{d^4k}{(2\pi)^4} \frac{e^{- i k
\cdot x}}{2 k_{+} k_{-} - k_\perp^{2} + i \epsilon} \left[g_{\mu
\nu} - \frac{n^{Weyl}_\mu k_\nu + n^{Weyl}_\nu
k_\mu}{k_{+} +  i \epsilon {\rm sgn}{k_{-}}}\right],
\label{LFWeylcanprop} 
\end{equation}
where the LF-Weyl gauge vector $n^{Weyl}\mu = (n^{Weyl}_{-} = 1,
n^{Weyl}_{+} = n^{Weyl}_i = 0)$ chooses the LF-Weyl gauge
$n^{Weyl}\cdot A = A_{+} = 0$. However when one LF front $x^{+}
= const.$ is fixed as a quantization surface, then the LF-Weyl
and the LC gauges are fundamentally different\footnote{For
example, when the periodic boundary conditions are imposed on
the gauge fields, the LC gauge has to be modified
\cite{KallRob1994}, while the LF-Weyl gauge remains 
unchanged.}, the gauge vector $n^{LC}_\mu$ lies on the
quantization surface while $n^{Weyl}_\nu$ is perpendicular to
this surface. Recently the LF-Weyl gauge has been successfully
implemented in the canonical formulation of the quantum
electrodynamics in the finite LF volume \cite{PrzNausKall}.  In
this model, the canonical DLCQ \cite{PauliBrodsky} analysis
has been carried separately for different subsystems of gauge
vector fields and charged fermion fields. This simplifying
property of the LF Weyl QED is rather a unique situation the LF
approach, contrary to the equal-time intuitions. We connect it
with the simple structure of electromagnetic currents, where no
derivatives of fermion fields occur. We expect that this simple
structure will not be spoiled by the infinite LF volume and in
this paper we decided to analyse the case of electrodynamics
with fermions. We leave the more involved cases of charged
scalar fields 
and the non-Abelian gauge
interactions for the future publications. \\ 
Our first aim is
to study the structure of the gauge field sector therefore we
will start the canonical LF quantization for the models where
the fermion currents are expressed by external arbitrary
currents. Because the LF Weyl gauge does not fix the gauge
symmetry completely, we expect that there will remain dynamical
non-physical gauge modes. In the canonical equal-time approach
\cite{Lazzizzera}, \cite{Burnel} such modes generate the
causal LM spurious poles in the propagators and introduce
positively non-definite Hilbert space of states.  We expect
that a similar scenario will also appear in the LF
quantization: the gauge field propagator will take the form
(\ref{LFWeylcanprop}) and the physical states will be chosen by
means of the Gauss law implemented as a weak condition on
states.\\ 
Our second aim is to formulate the LF perturbative Feynman
rules for the full QED. As the basic cross-check of the LF
formulation we take the formal equivalence of the S-matrix
elements calculated according to the LF Fynman rules with the
corresponding elements found within the equal-time approach.
We use the same functional method of proof, as in
\cite{Yan1972}, for all orders of perturbation in the coupling
constant $e$. \\
This paper is organized as follows. In Section 2 we analyze
the 1+1 dimensional model of the Abelian gauge fields
coupled linearly to external currents. The vector gauge field
propagators are given with a consistent definition for
naively coinciding spurious and physical poles. The higher
dimensional model is analysed in Section 3. The infra-red (IR)
singularities, encountered in the analysis of the independent
gauge modes, are dimensionally regularized and the final
expression for the gauge field propagator is found to 
have a finite limit in 3+1 dimensions. In Section 4
we reintroduce fermion fields and define the perturbative
Dyson theory in the interaction representation. The S-matrix
elements are proved to be (formally) equaivalent to those
calculated by the covariant (equal-time) Feynman rules. 
At the end all these results are discussed and the
future investigations are outlined. \\
All notations and definitions of Green functions are
given in Appendix A. In Appendix B the analytical and massive 
regularizations of IR singularities are presented in some
details. Appendix C contains calculations for the Poincar\'{e}
generators and their commutator relations.\\

\setcounter{equation}{0}

\section {Abelian gauge fields in 1+1 dimensions}

There are three reasons why we decided to start our analysis
with the simple 1+1 dimensional gauge field model. First,
inspecting the ML-pole in (\ref{LFWeylcanprop}) we find that it
contains only the longitudinal momenta $k_\pm$, which are
present already in 1+1 dimensions. Second, 
when we put $k_\perp \equiv 0$ in (\ref{LFWeylcanprop}), two
poles: physical and spurious coincide at $k_{+} = 0$. If the LF
canonical quantization will correctly reproduce these results,
then one can also expect its usefullness also for more physical
models in higher dimensions. 
Third, the Wilson loop calculations for the Yang-Mills
fields in $1+(D-1)$ dimensions show singularity at $D=2$
\cite{BassettoNardelli97}. Though we restrict ourselves to the 
Abelian fields, it may be instructive to compare gauge field
sectors in 1 + 1 and higher dimensions.\\
In this paper our analysis of the quantum electrodynamics in
the 1+1 dimensions will be limited to the canonical
quantization of Abelian vector gauge field coupled with the
external currents $j^\mu$. These currents describe couplings
with the fermion fields $J^\mu = - e \bar{\psi}\gamma^\mu
\psi$ when the fermion dynamics is abandoned. This means that
no conservation of $j^\mu$ is supposed, even in opposite, all
its components are treated as the arbitrary functions of
space-time. \\
On can impose  the Weyl gauge condition $A_{+} = 0$ explicitly
on the gauge fields $A_\mu$ and, as the starting point, one can
take the simple Lagrangian density 
\begin{equation}
{\cal L}^{1+1}_{Weyl}  =  \frac 1 2 \left(\partial_{+} {A}_{-}
\right)^2 + {A}_{-} {j}^{-}.
\end{equation} 
This reduced model has only one Euler-Lagrange
equation  
\begin{equation}
\partial_{+}^2{A}_{-} =   j^{-} \label{ELeq11},
\end{equation}
which evidently is a dynamical equation and one canonical
momentum $\Pi = \partial_{+} A_{-}$. We stress that there are
no primary constraints, but instead  we notice that the 
Gauss law 
\begin{equation}
G^{1+1} = \partial_{-} \partial_{+} A_{-} + j^{+} = \partial_{-} \Pi +
j^{+} = 0
\end{equation}
is lost. The canonical quantization is immediate and one
obtains the canonical Hamiltonian density 
\begin{equation}
{\cal H}^{1+1}_{can} =  \Pi \partial_{+} A_{-} - {\cal L}_{Weyl}^{1+1}
= \frac 1 2 (\Pi)^2 - A_{-} j^{-}\label{11canHam}
\end{equation}
and the nonvanishing canonical commutator
\begin{equation}
\left[ \Pi (x^{+}, x^{-}), A_{-}(x^{+},y^{-}) \right]  =  - i\ 
\delta(x^{-} - y^{-})\label{com11PiAm}.
\end{equation}
They generate the Hamilton equations of motion
\begin{equation}
\partial_{+}  \Pi  = j^{-} \ , \ \ \
\partial_{+} A_{-}  =  \Pi^{-}\ ,\label{1+1hameq}
\end{equation} 
which are evidently equivalent to the former Euler-Lagrange equation 
(\ref{ELeq11}). The Gauss law $G^{1+1}=0$ cannot be imposed
strongly as a 
condition on $\Pi$, because this would be inconsistent with
the above Hamiltonian structure. Therefore our quantum
theory describes a larger system than the physical
electrodynamics and one needs to impose the Gauss law rather as
a weak condition on physical states 
\begin{equation}
\langle phys'| G^{1+1}(x) |phys \rangle = 0
\label{11weakGauss}. 
\end{equation} 
The interaction part of (\ref{11canHam}) shows that there is no
instant current-current interaction\footnote{In the full
QED, the interaction Hamiltonian contains also dynamical
fermions $\psi_{+}$, $\psi_{+}^{\dag}$ and has a more 
complicated structure, however with no instantanous interaction
term. The perturbative theory based on the full interaction
Hamiltonian can be easily infered from the results given 
in the section {\ref{pertsec}} by omitting the transverse directions
and components.} quite contrary to the Hamiltonian in
the LC gauge \cite{tHooft1974}.  Therefore the perturbative
gauge field propagator is given just by the $x^{+}$-chronological
product of free gauge field operators $D_{--}(x,y) = i \langle 0
| T \ A_{-}({x}) A_{-}({y})|0\rangle$. When we put $j^{-} = 0$
in Eq. (\ref{1+1hameq}), the free field equations are immediately solved 
in terms of the $x^{+}$-independent operators $\pi(x^{-})$ and
$a_{-}(x^{-})$
\bml 
\begin{eqnarray}
\Pi(x) & = & \pi(x^{-}),\\
A_{-}(x) & = & x^{+} \pi(x^{-}) + a_{-}(x^{-}),
\end{eqnarray}
\eml
which have to satisfy the commutation relation  
\begin{equation}
\left [\pi({x^{-}}), a_{-}({y^{-}}) \right ]  =  - i \delta
(x^{-}-y^{-}).
\end{equation}
Introducing the Fourier representation for these free fields
\bml
\begin{eqnarray}
a_{-}(x^{-}) & = & \int_0^\infty \frac{dk_{-}}{2 \pi}
\ \left[  e^{+i k_{-} x^{-}} a^{\dag}(k_{-})+
+ e^{-i k_{-}x^{-}} a(k_{-}) + \right]\label{Fouram11}\\
\pi(x^{-}) & = &  \int_0^\infty \frac{dk_{-}}{2 \pi}
\ \left[ e^{+i k_{-} x^{-}} p^{\dag}(k_{-}) + e^{-i
k_{-}x^{-}} p(k_{-}) \right]\label{Fourpi11}
\end{eqnarray}
\eml
we get leads to the commutation relations for creation and annihilation
operators 
\bml
\begin{eqnarray}
\left[ a(k_{-}), p^{\dagger}(k'_{-})\right] & = &
2 i \pi \ \delta(k_{-} - k'_{-})\\
\left[ p(k_{-}), a^{\dagger}(k'_{-})\right] & = &
2 i \pi \ \delta({k_{-}} - {k'_{-}}).\label{11ampicom}
\end{eqnarray}
\eml
Because the above expressions show no singularity at the point $k_{-}
= k^{+} = 0$, we find that the propagating massless gauge
fields are less singular than the massless scalar fields in 1+1
dimensions \cite{Hagen1975}. Also we notice that these gauge fields
describe non-physical excitations because $\beta^{\dag}=
a^{\dag} + p^{\dag}$ operators can create Fock states
with negative metric. The Gupta-Bleuler method can be easily
applied here by imposing the weak Gauss law condition
(\ref{11weakGauss}) or equivalently
\begin{equation}
p (k_{-}) |phys \rangle = 0.
\end{equation}
Thus all physical states have zero norm and are created by 
the operator $p^{\dag}$ acting on the vaccun state $| 0
\rangle$.  Usually, in the equal-time models, the
zero-norm states are also present 
in the physical subspace while accompanying physical photon
states. In the LF picture, the physical photons are
described by transverse $A_\perp$ gauge fields and such fields are
evidently absent in 1+1 dimensions, so here we end up with the
physical subspace built solely from the zero-norm states.\\ 
In the large Hilbert space we immediately find propagators for the
independent free fields
\bml
\begin{eqnarray}
\langle 0 | T^{+} A_{-}(x) \Pi(y)|0\rangle & = &
\Theta(x^{+}-y^{+}) \langle 0 | a_{-}(x^{-}) \pi(y^{-})
|0\rangle + \Theta(y^{+}-x^{+}) \langle 0 | \pi(y^{-})
a_{-}(x^{-}) |0\rangle \nonumber\\
& = & E^1_F(x-y)\\
\langle 0 | T^{+} A_{-}({x}) A_{-}({y})|0\rangle 
& = & x^{+}\langle 0 | T\ \pi(x^{-}) a_{-}(y^{-})|0\rangle
+ y^{+} \langle 0 | T\ a_{-}(x^{-}) \pi(y^{-}) |0\rangle
\nonumber\\ 
&= & E^2_F(x-y),
\end{eqnarray}
\eml
where the noncovariant Green functions $E^1_F(x)$ and $E^2_F(x)$
are defined in Appendix A and we write a well
defined Fourier representation for the gauge field propagator
\begin{equation}
D_{--}(x) = i \langle 0 | T^{+} A_{-}({x}) A_{-}(0)|0\rangle = -
\int_{-\infty}^{\infty} \frac{d^2{k}}{(2 \pi)^2} e^{-i 
{k}\cdot x} \frac{1}{[k_{+} + i \epsilon \ {\rm
sgn}\ (k_{-})]^2} .
\end{equation}
Also we notice a property analogous to that for the LC gauge
propagators \cite{Bassetto1993}
\begin{equation}
D_{--}(x^{+}, x^{-}) =  2 \partial_{-} \int_{0}^{x^{+}} d\xi
D_F(\xi, x^{-}), 
\end{equation}
where $D_F(x)$ is the canonical light-front 
Feynman propagator for a massless scalar field \cite{Hagen1975}
\begin{equation}
D_F(x) = - \frac{1}{4\pi} \left[ \ln |x^{-}| + \frac{i\pi}{2}\ {\rm
sgn}(x^{+}) {\rm sgn}(x^{-}) \right] + f(x^{+}).
\end{equation}
Our conclusion is that in 1 + 1 dimensions only propagation of 
non-physical modes is allowed by the Weyl gauge condition and this
gives rise to a regular expression for the gauge field propagator.
Also we encounter no at $k_{-} = 0 $ - quite contrary to the 
expections for a 1 + 1 dimensional model with massless
excitations.

\setcounter{equation}{0}

\section{Canonical quantization in D dimensions}

In higher dimensions our model has new features, which were
absent in the previous case: the vector fields have their
transverse components $A_i$ and can can propagate also in the
transverse directions $x_\perp$. While the first property
introduces physical photons, then the second one will
lead to the unexpected IR singularities. Therefore
anticipating future problems, we choose to work in D dimensions,
where 2 coordinates are LF longitudinal $x_ L = (x^\pm)$ and
remaining $d = D-2$ coordinates are LF transverse
$x_\perp$\footnote{For notation see Appendix A.}. The canonical
LF quantization will be carried in $D>4$ dimensions, and the 
the limit $D \rightarrow 4$ will be taken for all expressions
which, being IR-finite, defint the perturbative QED in 3+1
dimensions.  Here again one can check that the canonical
quantization of the vector gauge fields can be done
independently from the fermion sector, therefore the fermion
contribution can be first decribed by the external currents
$j^\mu $. As we did in the previous section, we explicitly
implement the Weyl gauge condition $A_{+} = 0$ in the
Lagrangian density 
\begin{equation}
{\cal L}_{Weyl}  =  \partial_{+} {A}_i (
\partial_{-} {A}_i - \partial_i {A}_{-}) + \frac 1 2
\left(\partial_{+} {A}_{-} \right)^2 - \frac 1 4 \left(
\partial_i {A}_j- \partial_j {A}_i\right)^2 + {A}_{-} {j}^{-} +
{A}_i j^i .\label{AbelLagr} 
\end{equation}   
Our analysis begins with the Euler-Lagrange equations for
remaining vector gauge fields
\bml\label{3.2}
\begin{eqnarray}
\partial_{+}\left( \partial_{+}{A}_{-} - \partial_j A_j
\right) & = &  j^{-} ,\label{eqAm}\\ 
\left( 2\partial_{+} \partial_{-} - \Delta^d_\perp \right){A}_{i}
& = & \partial_{i}\left( \partial_{+}{A}_{-} - \partial_j A_j
\right) + {j}^{i} \label{eqAi},
\end{eqnarray}
\eml
where $\Delta^d_{\perp} = (\partial_j)^2 $ is the Laplace
operator in $d=D-2$ dimensions and again the Gauss law $G^{D} =
0$ is lost, where
\begin{equation}
G^{D} = \partial_{-}(\partial_{+} A_{-} - \partial_i A_i) -
\Delta_\perp^d A_{-} + j^{+}.
\end{equation}
Now we have the primary constraints (in Dirac's nomenclature)
connected with the canonical momenta $\Pi^i = 
\partial_{-} A_i - \partial_i A_{-}$, which   are not independent
canonical variables>. However these momenta cancel in the canonical
Hamiltonian density 
\begin{equation}
{\cal H}_{can}  =  \Pi^{-}\partial_{+}A_{-} +
\Pi^i\partial_{+}A_i - {\cal L}_{Weyl} 
= \frac 1 2 \Pi^2 + \frac 1 2 (\partial_i A_j)^2 + \Pi
\partial_i A_i  -   A_i J^i - A_{-} J^{-}\label{Hamden1}
\end{equation}
where $\Pi^{-} = \partial_{+}A_{-} = \Pi + \partial_i A_i$, so 
one can proceed freely without any further reference to the
constrained momenta $\Pi^i$.  The remaining fields: $ A_i, A_{-}$
and $\Pi$ are the independent canonical variables with their
equations of motion  
\bml\label{3.5}
\begin{eqnarray}
\left( 2\partial_{+} \partial_{-} - \Delta_\perp \right){A}_i
& = & \partial_i  \Pi  + j^i, \\
\partial_{+}  \Pi & = & j^{-},\\
\partial_{+} A_{-} & = & \Pi + \partial_i A_i,
\end{eqnarray} 
\eml
being equivalent to equations (\ref{3.2}) and the nonzero
equal-$x^{+}$ commutators
\bml
\begin{eqnarray}
\left [\Pi (x^{+},\vec{x}), A_{-}(x^{+},\vec{y}) \right] & = & -
i \delta^{d+1}(\vec{x} - \vec{y})\\
\left [2 \partial_{-} A_i (x^{+},\vec{x}), A_j(x^{+},\vec{y}) \right]
& = & - i \delta_{ij} \delta^{d+1}(\vec{x} - \vec{y})
\label{comAiPim} 
\end{eqnarray}
\eml
where $ \delta^{d+1}(\vec{x}) = \delta(x^{-})
\delta^{d}(x_\perp)$. The structure of dynamical equations
(\ref{3.5}) indicates our canonical variables are not yet 
the independent modes, which are very useful in the further
formulation. Therefore we propose to introduce the following
decomposition of vector fields 
\begin{equation}
A_{i} =  C_i - \partial_i [\Delta^d_\perp]^{-1}*(\Pi +
2 \partial_j C_j)\label{fromAitoCi} , \ \ \ 
A_{-}  =   C_{-} -  \partial_{-}[\Delta^d_\perp]^{-1}*(\Pi +
2 \partial_j C_j)\label{fromAmtoCm}
\end{equation}
which has the form of gauge transformation and its inverse 
from $\vec{C}$ to $\vec{A}$ has the same functional form.  The Green
function for the Laplace operator is well
defined for $d > 2$ 
\begin{equation}
[\Delta^d_\perp]^{-1}(x_\perp) = - \int \frac{d^dk_\perp}{(2\pi)^d}
\frac{e^{i k_\perp \cdot x_\perp}}{k_\perp^2} = -
\frac{1}{4 \pi} 
\frac{\Gamma(d/ 2 -1)}{(\pi
x_\perp^2)^{d/ 2-1}} 
\end{equation}
and the last analytical expressin is regular also for
noninteger $d < 2$. \\
There are also other regularizations of the inverse Laplace
operator strictly in $d=2$ dimensions \cite{Balasinetal1992}:
the analytical regularization - when the square pole is changed
to $2 -2 \delta$ where $ 1 > \delta > 0$ 
\begin{equation}
[\Delta_\perp]^{-1}(x_\perp) \rightarrow
[\Delta_{2 - 2\delta}]^{-1}(x_\perp) = (-1)^{1-\delta} \int
\frac{d^2k_\perp}{(2\pi)^2} \frac{e^{i k_\perp \cdot
x_\perp}}{(k_\perp^2)^{1-\delta}} = - \frac{1}{4\pi}
\frac{\Gamma(\delta)}{\Gamma (1-\delta)}
\left(-\frac{4}{x^2_\perp}\right)^\delta 
\end{equation}
and the massive regularization - when the pole at $k^2_\perp=0$
is shifted by the mass parameter $m^2$ 
\begin{equation}
[\Delta_\perp]^{-1}(x_\perp)  \rightarrow
[\Delta_\perp - m^2]^{-1}(x_\perp) = - \int \frac{d^2k_\perp}{(2\pi)^2}
\frac{e^{i k_\perp \cdot x_\perp}}{k_\perp^2+m^2} = - \frac{1}{2
\pi} K_{0}(m\sqrt{x_\perp^2}).
\end{equation}
However our decomposition formula (\ref{fromAitoCi}) cannot be
naively regularized in $d=2$ dimensions, because the analytical
regularization would bring different dimensions for two terms
with $C_i$, while the massive regularization would lead to the 
ill-defined inverse transformation (from $C_i$ to $A_i$). 
The consistent implementation of these two regularizations is
presented in Appendix \ref{Appreg}, while in this sectin we
will focus on the dimensional regularization.\\
For the independent modes one easily finds nonvanishing commutators 
\bml\label{3.11}
\begin{eqnarray}
\left [\Pi (x^{+},\vec{x}), C_{-}(x^{+},\vec{y}) \right] & = & -
i \delta^{d+1}(\vec{x} - \vec{y}), \label{DBPiCm}\\
\left [2 \partial_{-} C_i (x^{+},\vec{x}), C_j(x^{+},\vec{y}) \right]
& = & - i\delta_{ij} \delta^{d+1}(\vec{x} - \vec{y}),\label{DBCiCj}
\end{eqnarray}
\eml
and effective equations of motion 
\bml
\begin{eqnarray}
\left( 2\partial_{+} \partial_{-} - \Delta_\perp \right){C}_i
& = & j^i - 2 \partial_i [\Delta_\perp^d]^{-1}*(\partial_{-}j^{-}
+ \partial_j j^j) \label{HameqCi},\\ 
\partial_{+} C_{-} & = &- [\Delta_\perp^d]^{-1}* \left(
\partial_i j^i + \partial_{-} j^{-} \right)\label{HameqCm}\\ 
\partial_{+} \Pi & = & j^{-}.\label{HameqPi}
\end{eqnarray}
\eml
Also the Hamiltonian density (\ref{Hamden1}) can be
expressed in terms of these new fields 
\begin{equation}
{\cal H}_{eff}  =  \frac 1 2 (\partial_i C_j)^2 - \Pi
[\Delta_\perp^d]^{-1}* \left[\partial_i j^{i} + \partial_{-}
j^{-} \right] - C_i \left[j^i - 2\partial_i
[\Delta_\perp^d]^{-1}*(\partial_{-}j^{-} + \partial_j 
j^j) \right] - C_{-} j^{-},
\end{equation}
showing again no instantaneous current-current interaction.
Therefore, in the interaction picture, the perturbative
propagators for independent modes will be given by the
$x^{+}$-chronological products of free fields. 
The Fourier representation of free fields is defined  for
positive values of the $k_{-}$ momentum 
\bml
\begin{eqnarray}
C_{-}(\vec{x}) & = & \int_{-\infty}^{\infty} \frac{d^dk_\perp}{(2
\pi)^d} \int_{0}^{\infty} \frac{dk_{-}}{2\pi}
 \ \left[e^{-i \vec{k}\cdot
\vec{x}} a(\vec{k}) + e^{+i \vec{k}\cdot \vec{x}}
a^{\dagger}(\vec{k})\right]\\
\Pi(\vec{x}) & = &\int_{-\infty}^{\infty} \frac{d^dk_\perp}{(2
\pi)^d} \int_{0}^{\infty} \frac{dk_{-}}{2\pi}
 \ \left[e^{-i \vec{k}\cdot
\vec{x}} p(\vec{k}) + e^{+i \vec{k}\cdot \vec{x}}
p^{\dagger}(\vec{k})\right]\\
C_{i}(x) & = & \int_{-\infty}^{\infty} \frac{d^dk_\perp}{(2
\pi)^d} \int_{0}^{\infty} \frac{dk_{-}}{2\pi\ 2k_{-}}
 \ \left[e^{-i {k}\cdot{x}} c_i(\vec{k}) + e^{+i
{k}\cdot {x}} c^{\dagger}_i(\vec{k})\right]_{k_{+} =
\frac{k_\perp^2}{2k_{-}}},
\end{eqnarray}
\eml
and the commutators for creation and annihilation operators
follow from (\ref{3.11})
\bml
\begin{eqnarray}
\left[ a(\vec{k}), p^{\dagger}(\vec{k'})\right] & = & 
\left[ a^{\dagger}(\vec{k}), p(\vec{k'})\right] = i (2\pi)^{d+1}
\delta^{d+1}(\vec{k} - \vec{k'})\label{comrelcp}\\ 
\left[ c_{i}(\vec{k}), c^{\dagger}_j(\vec{k'})\right] & =
&(2\pi)^{d+1} \ 2 k_{-} \ \delta_{ij}\delta^{d+1}(\vec{k} -
\vec{k'})
\end{eqnarray}
\eml
where $\delta^{d+1}(\vec{k}) = \delta^d(k_\perp) 
\delta(k_{-})$\footnote{The commutators (\ref{comrelcp}) show that the
operators $a^{\dag}$ and ${p}^{\dag}$ are trivial
generalizations of the respective operators in the 1+1
dimensions and therefore here again they can create negative
norm Fock states and the physical subspace has to be chosen by
means of the weak Gauss law.}. The nonzero propagators for the
independent modes can be easily found 
\bml
\begin{eqnarray}
\langle 0 | T^{+}\ C_{-}(\vec{x}) \Pi(\vec{y})|0\rangle & = &
E_F^{1}(x_L-y_L)\delta^d(x_\perp - y_\perp) ,\\
\langle 0 | T^{+}\ C_{i}({x}) C_j({y})|0\rangle & = &
\delta_{ij} D_F^{d+2}(x-y),
\end{eqnarray}
\eml
where the covariant Feynman Green function $D^{d+2}_F(x)$ is
defined in Appendix \ref{AppGreen}. Next, one can use the
decompositions (\ref{fromAitoCi}) when calculationf the
propagators for the fields $\vec{A}$. For the transverse
components $A_i$ only the contribution of $C_i$ fields is taken
\bml
\begin{eqnarray}
\langle 0 | T^{+} A_{i}({x}) A_j({y})|0\rangle & = & \left(\delta_{ik} -
2 \partial_i^x \partial_k^x [\Delta^d_\perp]^{-1}
*\right)\left(\delta_{jl} - 2 \partial_j^y\partial_l^y
[\Delta^d_\perp]^{-1} * \right) D_F^{d+2} (x - y) \delta_{kl}\nonumber\\
 &=& 
\delta_{ij} D_F^{d+2}(x-y),
\end{eqnarray}
but for the longitudinal component $A_{-}$ there are more terms
\begin{eqnarray}
\langle 0 | T^{+}  A_{-}({x}) A_j({y})|0\rangle & = &
\partial_j^x [\Delta_\perp^d]^{-1}* \left\{ 2 \partial_{-}^x
D_F^{d+2}(x-y) + E_F^{1}({x}_L-{y}_L)\delta^d(x_\perp-y_\perp) \right\} \\
\langle 0 | T^{+} A_{-}({x}) A_{-}({y})|0\rangle & = &
2 \partial_{-}^x [\Delta_\perp^d]^{-1}* \left\{2 \partial_{-}^x
D_F^{d+2}(x-y) + E_F^{1}({x}_L-{y}_L)\delta^d(x_\perp-y_\perp)\right\}. 
\end{eqnarray}
\eml
The expression in the above curly brackets can be expressed either as the
integral over $x^{+}$ of the covariant Feynman Green function
\bml\label{3.30}
\begin{eqnarray}
\langle 0 | T^{+} A_{-}({x}) A_{i}({y})| 0\rangle 
&=& \partial_j \int_0^{x^{+}-y^{+}} dz^{+} D^{d+2}_F(z^{+} +
y^{+}, \vec{x}-\vec{y})\ , \\ 
\langle 0 | T^{+} A_{-}({x}) A_{-}({y}) |0\rangle  
& = & 2 \partial_{-}\int_0^{x^{+}-y^{+}} d\xi D^{d+2}_F(z^{+} +
y^{+}, \vec{x}-\vec{y})
\end{eqnarray}
\eml
or as the Fourier integral with two causal poles 
\begin{equation}
\langle 0 | T^{+} A_{-}({x}) A_{\nu}({y})|0\rangle = 
i \int \frac{d^{d+2} k}{(2 \pi)^{d+2}} \frac{e^{-i {k}\cdot
({x}-{y})}}{k^2 + i \epsilon} \frac{(k_{-} n^{Weyl}_\nu + k_\nu
n^{Weyl}_{-})}{k_{+} + i\ \epsilon' \ {\rm sgn}(k_{-})}.
\end{equation}
Finally the Fourier representation can be written in a concise
form for all components of $A_\mu$ field
\begin{equation}
\langle 0 | T^{+} A_{\mu}({x}) A_{\nu}({y})|0\rangle = 
i \int \frac{d^{d+2} k}{(2 \pi)^{d+2}} \frac{e^{-i {k}\cdot
({x}-{y})}}{k^2 + i \epsilon} \left[- g_{\mu \nu}
+ \frac{(k_\mu n^{Weyl}_\nu + k_\nu n_\mu^{Weyl})}{k_{+}
+ i\ \epsilon' \ {\rm sgn}(k_{-})} \right].
\end{equation}
All expressions for propagators have the limit $d \rightarrow 2$
and so we arrive at the canonical LF propagators in $3+1$
dimensions 
\begin{equation} 
\langle 0 | T^{+} A_{\mu}({x}) A_{\nu}({y})|0\rangle =  i \int
\frac{d^{4} k}{(2 \pi)^{4}} \frac{e^{-i {k}\cdot
({x}-{y})}}{k^2 + i \epsilon} \left[-g_{\mu \nu} + \frac{(k_\mu
n^{Weyl}_\nu + k_\nu n^{Weyl}_\mu)}{k _{+} + i\ \epsilon' \
{\rm sgn}(k_{-})} \right]\label{finalpropagator} .
\end{equation}
The relations (\ref{3.30}), which are evidently valid also in $3+1$
dimensions, can be shown to be the zeroth order (in powers of
e) terms of the integrated Schwinger-Dyson relations for the
full QED in the presence of interactions\footnote{These
problems will be discussed in a future publication.}. \\
Our above canonical LF analysis gives the same result for the gauge
field propagator as the equal-time quantization. In both
approaches the non-physical modes are allowed to propagate and
they introduce causal ML prescription for axial spurious
poles.\\ 
Before giong to the interacting model with fermions, let us
study the Poincar\'{e} covariance in the gauge field sector. We
notice that both the choice of a null-surface $x^{+} = const$
as the quantization surface and by the noncovariant LF-Weyl
gauge condition $A_{+}= 0$ violate the Poincar\'{e} covariance.
This is quite similar to the equal-time canonical procedure
for the temporal gauge condition $A_{0} = 0$ \cite{Lazzizzera},
where the Poincar\'{e} covariance is recovered in the physical
subspace selected by the weak Gauss law. In our LF formulation
we will check this possibility in 3 + 1 dimensions.\footnote{In
Appendix D we give the definitions of Poincar\'e generators and
present some nontrivial steps of calculations.} 
First we check that $\vec{A}$ transforms covariantly
\begin{equation}
\left[M^{\mu \nu}, A^\lambda(x)\right] = -
i\left(x^{\mu}\partial^\nu - x^\nu \partial^{\mu}\right)
A^\lambda(x) - i\left(g^{\mu \lambda} A^\nu - g^{\nu \lambda}
A^\mu\right)(x). 
\end{equation}  
Then we analyse the commutator algebra of canonical Poincar\'{e}
generators 
\begin{eqnarray}
\left[ M^{\mu \nu}, P^{\lambda}\right] &  = & i g^{\nu \lambda} P^\mu
- i g^{\mu \lambda} P^\nu + i \int d^3\vec{x}
\delta^{\lambda -} \left[ g^{\mu -} G(x)A^\nu(x) - g^{\nu -}
G(x)A^{\mu}(x) \right]_{sym} \label{PoinccomMP}\\
\left[M^{\mu \nu}, M^{\lambda \rho}\right] & = & 
+ i \left( g^{\mu \lambda} M^{\nu \rho}- g^{\nu \lambda} M^{\mu
\rho} - g^{\mu \rho} M^{\lambda\nu }- g^{\nu \rho} M^{\lambda
\mu}\right) \nonumber\\ 
& + & i \int d^3\vec{x} \left( x^\mu g^{\nu +} - x^\nu
g^{\mu +}\right) \left[ g^{\lambda -}G(x)A^\rho(x) - g^{\rho -}
G(x)A^{\lambda}(x) \right]_{sym}\nonumber\\
& - & i \int d^3\vec{x} \left( x^\lambda g^{\rho +} - x^\rho
g^{\lambda +}\right) \left[ g^{\mu -}G(x)A^\nu(x) - g^{\nu -}
G(x)A^{\mu}(x) \right]_{sym}\label{PoinccomMM}
\end{eqnarray}
where $G = \partial_{-}(\partial_{+} A_{-} +  \partial_i A_i) - \Delta_\perp
A_{-}$ is the Gauss law operator in the absence of charged
currents and the symmetrization for noncommuting $G$ and $A^\mu$
operators is imposed. In the physical subspace with states
selected by the weak Gauss law $\left \langle phys' | G(x) |
phys \right \rangle = 0$
one finds that the anomalous terms  vanish 
\begin{equation}
\left \langle phys' \left| \left[ g^{\mu -}G(x)A^\nu(x) - g^{\nu -}G(x)
A^{\rho}(x) \right]_{sym}\right| phys \right \rangle  = 0
\end{equation}
and one concludes that the Poincar\'{e} covariance is recovered
here.\\

\setcounter{equation}{0}

\section{Interacting theory with fermions}

If one takes the effective Lagrangian density for the gauge
sector in the form
\begin{equation}
{\cal L}_{gauge}^{eff} = \Pi \partial_{+} A_{-} + \partial_{-}A_i
\partial_{+} A_i - \frac 1 2 \Pi^2 - \frac 1 2 (\partial_i
A_j)^2 - \Pi \partial_i A_i + A_i j^i + A_{-}j^{-}\label{effLagr31}, 
\end{equation}
then one can incorporate fermions by resubstituting $j^{\mu}$
by the fermion currents $- e\bar{\psi} \gamma^\mu
\partial_{\mu} \psi$ and adding  the 
kinetic terms for fermions. All this leads to the following
Lagrangian density for the interacting theory
\bml
\begin{eqnarray}
{\cal L}_{Weyl}^{QED} & = &
\Pi \partial_{+} A_{-} + \partial_{-}A_i
\partial_{+} A_i - \frac 1 2 \Pi^2 - \frac 1 2 (\partial_i
A_j)^2 - \Pi \partial_i A_i + i \sqrt{2} \psi^{\dag}_{+} \partial_{+} 
\psi_{+} \nonumber\\
& + & \sqrt{2} \psi^{\dag}_{-} \left( i \partial_{-} - e
A_{-}\right)\psi_{-} 
+ \xi^{\dag} \psi_{-} + \psi_{-}^{\dag}\xi \label{LagreffQED}
\end{eqnarray}
where $\psi_{\pm} = \Lambda_{\pm} \psi$, $\psi_{\pm}^{\dag}  =
 \psi^{\dag} \Lambda_{\pm}$ and  
\begin{eqnarray}
\xi & = & \left[- i \partial_i \alpha^i + M \beta 
+ e A_i  \alpha^i \right]\psi_{+}\ , \ 
\xi^{\dag}  =  \psi^{\dag}_{+} \left[ i \stackrel{\leftarrow}{\partial_i} \alpha^i +
M \beta  + e  \alpha^i A_i\right] .\label{defxis}
\end{eqnarray}
\eml
As usually in the LF
formulation, the fermions $\psi_{-}$ and $\psi^{\dag}_{-}$
fermions satisfy nondynamical nondynamical equations
\begin{equation}
\sqrt{2}\left(i\partial_{-} - e A_{-}\right) \psi_{-}  = \xi\ , \ \
- \sqrt{2}\left(i\partial_{-} + e A_{-}\right)\psi_{-}^{\dag}
=  \xi^{\dag} . \label{nondynfermeq}
\end{equation}
For solving these equation uniquely, one has to impose some
boundary conditions for the dependent fermion fields. Here we
choose the simplest and commonly used possibility - the
antisymmetric conditions
\begin{equation}
\lim_{x^{-} \to - \infty} \psi_{-}(x) = - \lim_{x^{-} \to 
\infty} \psi_{-}(x) , \ \ \ \ 
\lim_{x^{-} \to - \infty} \psi^{\dag}_{-}(x) = - \lim_{x^{-}
\to  \infty} \psi_{-}^{\dag}(x) , 
\end{equation}
and this allows us to solve Eqs. (\ref{nondynfermeq}) as
\bml
\begin{eqnarray}
\psi_{-}(x) & = & \frac{1}{\sqrt{2}} \int_{-\infty}^{\infty} dy^{-}
\ \left(i\partial_{-} - eA_{-}\right)^{-1}[x^{-}, y^{-};
x_\perp] \xi(y^{-}, x_\perp), \\ 
\psi^{\dag}_{-}(x) & = & \frac{1}{\sqrt{2}}
\int_{-\infty}^{\infty} dy^{-} \ \left(i\partial_{-} -
eA_{-}\right)^{-1}[y^{-},x^{-}; x_\perp]
\xi^{\dag}(y^{-}, x_\perp), 
\end{eqnarray}
\eml
where the Green function $\left(i\partial_{-} - eA_{-}
\right)^{-1} [x^{-}, y^{-}, x_\perp]$ is defined by the 
integral equations 
\bml
\begin{eqnarray}
\left(i\partial_{-} - eA_{-}\right)^{-1}[x^{-}, y^{-}; x_\perp]
& = & \left(i\partial_{-}\right)^{-1}(x^{-} - y^{-}) +
\int^{\infty}_{-\infty} dz^{-} \left(i\partial_{-}\right)^{-1}
(x^{-} - z^{-}) \ eA_{-}(z^{-}, x_\perp)\nonumber\\
&& \hspace{110pt}\times \ \left(i\partial_{-} - 
eA_{-}\right)^{-1} [z^{-}, y^{-}; x_\perp]\\
& = &  \left(i\partial_{-} \right)^{-1}(x^{-} - y^{-}) +
\int^{\infty}_{-\infty} dz^{-} \left(i\partial_{-} -
eA_{-}\right)^{-1}[x^{-}, z^{-}; x_\perp] \nonumber\\ 
&& \hspace{110pt}\times \  eA_{-}(z^{-},
x_\perp) \left(i\partial_{-}\right)^{-1}
(z^{-}- y^{-}). 
\end{eqnarray}
\eml
It is useful to introduce the matrix notation for the integral
operators, by leaving the explicite dependence on coordinates, 
denoting the integration over $x^{-}$ variables by
asterisks and writing the integral operators as fractions; for
example the above equations are written as 
\begin{equation}
\frac{1}{i\partial_{-} - eA_{-}} =  \frac{1}{i\partial_{-}}  +
\left(\frac{1}{i\partial_{-}} eA_{-}\right) \ast 
\frac{1}{i\partial_{-} - eA_{-}}  =  \frac{1}{i\partial_{-}}  +
\frac{1}{i\partial_{-}- e A_{-}}\ast \left( eA_{-}
\frac{1}{i\partial_{-}}\right).
\end{equation}
These equations can be used for generating the series in the
arbitrary powers of e, provided the following integral
\begin{eqnarray}
\int^{\infty}_{-\infty} dz^{-}
\left(i\partial_{-}\right)^{-1}(x^{-} - z^{-}) \ eA_{-}(z,
x_\perp) \ \left(i\partial_{-} \right)^{-1} (z^{-}- y^{-})
\sim\nonumber\\ 
\sim \int^{\infty}_{-\infty} dp_{-} dk_{-}\ e^{-i p_{-} (x^{-}
- y^{-})} \widetilde{A}_{-}(k_{-}, x_\perp)\ {\rm CPV}\left[
\frac{1}{p_{-}}\right]\ {\rm CPV}\left[\frac{1}{p_{-}-k_{-}}
\right]
\label{expression}
\end{eqnarray}
is a well-defined expression. This means that one has to
consider $A_{-}$ fields with their momenta $k_{-} \neq 0$,
otherwise in Eq. (\ref{expression}) one would encounter
ill-defined expression $\displaystyle \left(\text{CPV}\left[
\frac{1}{p_{-}}\right]\right)^2$. Bearing this limitation in
mind we can write the infinite series for $\left(i\partial_{-}
- eA_{-} \right)^{-1}$ 
\bml
\begin{equation}
\frac{1}{i\partial_{-} - eA_{-}}  =  \frac{1}{i\partial_{-}}\ast
{\cal W}_{-1}[\widehat{a}] = {\cal W}_{-1}[\widehat{a}^{\dag}]
\ast \frac{1}{i\partial_{-}}
\end{equation}
where
\begin{eqnarray}
{\cal W}_{-1}[\widehat{a}] & = & \sum_{n = 0}^{n= \infty} 
\underbrace{\left(\frac{1}{i\partial_{-}} eA_{-}\right) \ast \ldots
\ast \left(\frac{1}{i\partial_{-}} eA_{-}\right)}_{n}, \\
{\cal W}_{-1}[\widehat{a}^{\dag}] & = & \sum_{n = 0}^{n= \infty}
\underbrace{\left(eA_{-} \frac{1}{i\partial_{-}} \right) \ast \ldots
\ast \left( eA_{-} \frac{1}{i\partial_{-}}\right)}_{n}.
\end{eqnarray}
\eml
Thus we write the fermion contribution to the Hamiltonian as
\begin{equation}
H_{fer} = \frac{1}{\sqrt{2}} \int_{-\infty}^{\infty} d^dx_\perp
\ \xi^{\dag} \ast \frac{1}{i\partial_{-}}\ast
{\cal W}_{-1}[\widehat{a}] \ast \xi,
\end{equation}
where the integrand is a local expression in the $x_\perp$
coordinates, and our notation treats $\xi$ as a column 
matrix and $\xi^{\dag}$ as a row matrix. \\
Thus we end up with the total Hamiltonian, which depends solely
on dynamical field variables 
\begin{eqnarray}
{H}_{total} & = & \int d^{d+1}\vec{x} \left[ \frac 1 2 \Pi^2 +
\frac 1 2 (\partial_i A_j)^2 + \Pi \partial_i A_i\right]
 + \frac{1}{\sqrt{2}} \int_{-\infty}^{\infty} d^dx_\perp
\ \xi^{\dag} \ast \frac{1}{i\partial_{-}}\ast
{\cal W}_{-1}[\widehat{a}] \ast \xi,  \label{totalHamilt}
\end{eqnarray}
while the nonvanishing equal $x^{+}$ (anti)commutators have the 
canonical LF form 
\bml\label{4.12}
\begin{eqnarray}
\left [\Pi (x^{+},\vec{x}), A_{-}(x^{+},\vec{y}) \right] & = & -
i \delta^{d+1}(\vec{x} - \vec{y})\label{31PiAmcommut},\\
\left [2 \partial_{-} A_i (x^{+},\vec{x}), A_j(x^{+},\vec{y}) \right]
& = & - i \delta_{ij} \delta^{d+1}(\vec{x}
-\vec{y}),\label{31Aiajcommut}\\ 
\left \{ \psi^{\dag}_{+}(\vec{x}), \psi_{+}(\vec{y}) \right \} & =
& \frac{1}{\sqrt{2}} \delta^3(\vec{x} - \vec{y}).
\label{31psipsidagantcom} 
\end{eqnarray}
\eml
Thus in the LF Weyl gauge $A_{+} = 0$ the quantum
electrodynamics of massive fermion matter fields contains no
instantanous interactions of currents and have all
(anti)commutator relations (at equal-$x^{+}$) independent of
interactions and c-numbered. All these properties, while being
quite normal in the equal-time formulation, are not such in the
front description of interacting theories, when the light-front
surface probes the fields at the light-like seperated points of
space-time. \\

\subsection{Perturbation theory}\label{pertsec}

The perturbation theory is defined most easily in the
interaction picture, where all quantum field operators have
free dynamics and the interaction appears in the evolution of
quantum states. Thus we also choose to work in this
representation, however for clarity of further formulas we will
omit the usually used subscript {\scriptsize I}. Free evolution
follows from a free Hamiltonian, but we notice that the total
Hamiltonian (\ref{totalHamilt}) depends on the 
coupling constant e both non-locally in the expression
${\cal W}_{-1}[\widehat{a}]$ and locally in the fields $xi$ and
$\xi^{\dag}$ - compare Eq.(\ref{defxis}). Thus we define the
free Hamiltonian $H_0$ as the limit 
\begin{eqnarray}
{H}_{0} &\stackrel{df}{=}& \lim_{e \to 0} {H}_{total} = 
\int d^{d+1}\vec{x} \left[\frac 1 2 \Pi^2 + \frac 1 2
(\partial_i A_j)^2 + \Pi \partial_i A_i\right] +
\int d^d x_{\perp} \ \frac{1}{\sqrt{2}} \xi^{\dag}_0 
\frac{1}{i \partial_{-} }* \xi_0 \ ,\label{freeHamilt}
\end{eqnarray}
where now we have 
\begin{equation}
\xi_0 = \left[- i \partial_i \alpha^i + M \beta \right]
\psi_{+}, \ \ \xi^{\dag}_0 = \psi^{\dag}_{+} \left [ i
\stackrel{\leftarrow}{\partial_i} \alpha^i + M \beta \right].
\end{equation}
Thus from the (anti)commutator relations (\ref{4.12}), which
being independent of interactions, remain unchanged, we find
the  free evolution equations
\bml
\begin{eqnarray}
(2 \partial_{+} \partial_{-} - \Delta_\perp) {A}_{i} & = &
\partial_i \Pi \ ,\\
\partial_{+} \Pi & = & 0 \ , \\
\partial_{+} A_{-} & = & \Pi + \partial_i A_i\ , \\
\left( 2 \partial_{+} \partial_{-} + M^2\right) \psi_{+} & = &
0 \ , \\ 
\left( 2 \partial_{+} \partial_{-} + M^2\right) \psi_{+}^{\dag}
& = & 0 .
\end{eqnarray}
\eml
The perturbative propagators are defined as the chronological
products of dynamical free fields
\bml
\begin{eqnarray}
\left\langle 0 \left| T^{+} A_\mu(x) A_\nu(y)\right| 0
\right\rangle  & = &  i
g_{\mu \nu} D_F(x-y) + (n_\mu \partial_{\nu} + n_{\nu}
\partial_{\mu} ) \left(E^1_F \ast D^{d+2}_F\right) (x-y)\ , \\
\left\langle 0 \left| T^{+} \psi_{+}(x) \psi_{+}^{\dag}(y)\right| 0
\right \rangle & = & i \sqrt{2} \Lambda_{+} \partial_{-}^x
\Delta_F^{d+2}(x-y, M^2)\ ,
\end{eqnarray}
\eml
where for the gauge fields we have repeated our previous
analysis, while the result for dynamical fermions is well 
known \cite{Yan1972}. These expressions will produce all
nontrivial contractions in the LF version of the Dyson
perturbation theory, while the vertices are to be 
taken from the interaction Hamiltonian, which is defined as the
difference between the total and the free Hamiltonians $
{H}_{int} = {H}_{total} - {H}_{int}$. Because all these
Hamiltonians are local in the transverse directions we can
write the expressions for the density (in $x_\perp$)  
\begin{eqnarray}
\widetilde{\cal H}_{int} & = & \frac{1}{\sqrt{2}} \xi^{\dag}_0
\ast \frac{1}{i \partial_{-}} \ast \left( eA_{-}
\frac{1}{i\partial_{-}} \right) \ast {\cal W}_{-1}[\widehat{a}]
\ast \xi_0 + \frac{e}{\sqrt{2}} \left(\psi^{\dag}_{+} \alpha^i
A_i\right) \ast \frac{1}{i \partial_{-}} \ast {\cal
W}_{-1}[\widehat{a}] \ast  \xi_0 \nonumber\\
&+& \frac{e}{\sqrt{2}} \xi^{\dag}_{0} \ast \frac{1}{i
\partial_{-}} \ast {\cal W}_{-1}[\widehat{a}] \ast \left(
\alpha^i A_i \psi_{+}\right) + \frac{e^2}{\sqrt{2}} \left(
\psi^{\dag}_{+} \alpha^j A_j \right)\ast \frac{1}{i
\partial_{-}} \ast {\cal W}_{-1}[\widehat{a}] \ast \left(
\alpha^i A_i \psi_{+} \right) \nonumber\\
&&\label{intHam1}\ . 
\end{eqnarray}
This expression looks fairly complicated, but fortunately, for
our choice of antisymmetric boundary conditions for
$(i\partial_{-})^{-1}$, we may consistently introduce the
fields $\Psi_{-}$ and $\Psi_{-}^{\dag}$ 
\begin{equation}
\Psi_{-}  =  \frac{1}{\sqrt{2}} \frac{1}{i \partial_{-}}*
\left[- i \partial_i \alpha^i - M \beta \right]\psi_{+} , \ \
\Psi^{\dag}_{-} = \psi^{\dag}_{+} \frac{1}{\sqrt{2}} \left[ i
\stackrel{\leftarrow}{\partial_i}  \alpha^i + M \beta \right]
*\frac{1}{i \partial_{-}},
\end{equation}
and further the chronological products for them
\bml
\begin{eqnarray}
&\left\langle 0 \left| T^{+} \Psi_{-}(x) \psi_{+}^{\dag}(y)
\right| 0 \right\rangle + \left\langle 0 \left| T^{+}
\psi_{+}(x) \Psi_{-}^{\dag}(y)\right| 0 \right\rangle = \left(
- \partial_i^x \gamma^i + i M\right) 
\gamma^0 \Delta_F^{d+2}(x-y, M^2)\\
&\left\langle 0 \left| T \Psi_{-}(x) \Psi_{-}^{\dag}(y)\right| 0
\right \rangle = - \partial_{+}^x \gamma^{+} \gamma^0
\Delta_F^{d+2}(x-y, M^2) - \gamma^{+} \gamma^0 \displaystyle
\frac{1}{2 \partial_{-}}* \delta(x-y). 
\end{eqnarray}
\eml
Now it is easy to check that for the new fermion fields 
$\Psi = \psi_{+} + \Psi_{-}$ and $\Psi^{\dag} =
\psi^{\dag}_{+}+ \Psi^{\dag}_{-}$ we have a compact expression
for the propagator
 \begin{equation}
\left \langle 0 \left| T \Psi(x) \bar{\Psi}(y)\right| 0
\right \rangle = - \left(\partial_{\mu }^x \gamma^{\mu} + i M
\right) \Delta_F(x-y, M^2) - \gamma^{+} \frac{1}{2
\partial_{-}}* \delta(x-y), \label{psibarpsiprop}
\end{equation}
while a more tedious algebra for the interaction Hamiltonian
(\ref{intHam1}) density gives 
\begin{equation}
\widetilde{\cal H}_{int} = e \bar{\Psi}  \left(\gamma^{-}
{A}_{-} + \gamma^{i} {A}_{i}\right) \ast \left[1 + \frac{e}{2}
\gamma^{+} \frac{1}{i \partial_{-}} \ast {\cal
W}_{-1}[\widehat{a}]  \left(
\gamma^{-} {A}_{-} + \gamma^{j} {A}_{j}\right)\right]\ast \Psi.
\label{modintHam1}. 
\end{equation}
These expressions can be compared to those obtained in the LC
gauge $A_{-}= 0$. In both cases the chronological product
for free fermion fields is the same (\ref{psibarpsiprop}) and
has one non-covariant non-causal term. Our interaction
Hamiltonian (\ref{modintHam1}) has the infinite number of
non-covariant vertices which are generated by ${\cal
W}_{-1}[\widehat{a}]$, contrary to the LC gauge Hamiltonian,
where there is only one noncovariant term. In the LC gauge
case, these two non-covariant terms: from the fermion progator
and from the interaction Hamiltonian, cancel each other
\cite{Yan1972}. A similar cancelation should also
appear in the LF Weyl gauge. However, due to the infinite
number of non-covariant terms in (\ref{modintHam1}), it is an
open question if this cancelation will be complete. 

\subsection{Structure of the S-matrix in the LF perturbative calculations}

Above we have noted that all non-covariant terms are connected
with the fermion fields, so in order to 
prove the equivalence of the LF perturbative theory and the
usual equal-time formulation we need to study only fermion
contractions. Keeping the gauge fields $A_\mu$ as c-numbers,
we study Wick's theorem for the chronological product 
\bml 
\begin{equation}
S_{fer}[\bar{\Psi}, \Psi; \vec{A}] = T^{+} \exp \left( - i e
\int d^dx_\perp \ \bar{\Psi} \ast {\cal V}\ast \Psi\right),
\end{equation}
where
\begin{equation}
{\cal V} = e \left(\gamma^{-} {A}_{-} + \gamma^{i} {A}_{i}\right)
\left[1 + \frac{e}{2} \gamma^{+} \frac{1}{i \partial_{-}} \ast
{\cal W}_{-1}[\widehat{a}] \left(\gamma^{-} {A}_{-} +
\gamma^{j} {A}_{j}\right)\right] .
\end{equation}
\eml
Using the Schwinger functional technique \cite{SchwingerFunct},
we derive theone gets the result 
\begin{eqnarray}
S_{fer}[\bar{\Psi}, \Psi; \vec{A}] & = & {\exp}\left[{\text Tr} \ln
\left(1 - \bar{S}_{F} \ast {\cal V} \right)\right] :
\exp  \ \left[- i e
\int d^dx_\perp \bar{\Psi} \ast {\cal V}(1 -
\bar{S}_{F} \ast {\cal V})^{-1}\ast  \Psi\right]:\ ,
\end{eqnarray}
where the normal ordering is taken only for fermion fields and 
\begin{equation}
 \bar{S}_F(x-y)  =  - i\left\langle 0 \left| T^{+} \Psi(x)
\bar{\Psi}(y)\right| 0 \right\rangle = \left( i\gamma^\mu
\partial_\mu^{x} - M\right) \Delta_F(x-y,M^2) - \frac{\gamma^{+}}{2}
\frac{1}{i\partial_{-}}* \delta(x-y).
\end{equation}
The algebra of Dirac matrices simplifies the contribution
for the non-covariant part of the fermion propagator 
\bml
\begin{equation}
- \frac{\gamma^{+}}{2}\frac{1}{i\partial_{-}}*{\cal V} = 
- \frac{\gamma^{+}}{2}\frac{1}{i\partial_{-}}*{\cal
W}_{-1}[\widehat{a}]\left(\gamma^{-} {A}_{-} +
\gamma^{j} {A}_{j}\right) ,
\end{equation}
and then one easily checks the following factorization property
\begin{equation}
1 - \bar{S}_{F} \ast {\cal V} = \left[1 - S^{cov}_F
\left(\gamma^{-} {A}_{-} + \gamma^{j} {A}_{j}\right) \right]
\ast \left[1 + \frac{e}{2} \gamma^{+} \frac{1}{i \partial_{-}} \ast
{\cal W}_{-1}[\widehat{a}] \left(\gamma^{-} {A}_{-} +
\gamma^{k} {A}_{k}\right)\right],
\end{equation}
\eml
where $S^{cov}_F$ is the covariant part of the fermion
propagator. It is also useful to analyse the non-covariant factor
\begin{equation}
\text{Tr} \ln \left[1 + \frac{e}{2} \gamma^{+} \frac{1}{i
\partial_{-}} \ast {\cal W}_{-1}[\widehat{a}] \left(\gamma^{-}
{A}_{-} + \gamma^{k} {A}_{k}\right)\right]= \frac 1 4 {\text tr} 1
\text{Tr} \ln {\cal W}_{-1}[\widehat{a}^{\dag}]. 
\end{equation}
Gathering all above results we conclude that almost all
non-covariant terms have canceled
\begin{eqnarray}
S_{fer}[\bar{\Psi}, \Psi; \vec{A}] & = & : {\exp} \left[ - i e
\int d^dx_\perp \bar{\Psi} \left(\gamma^{-} {A}_{-} +
\gamma^{k} {A}_{k}\right)\ast \left[1 - {S}^{cov}_{F}
\ast \left(\gamma^{-} {A}_{-} + \gamma^{k}
{A}_{k}\right)\right]^{-1}\ast  \Psi\right] :
\nonumber\\
&&  {\exp}\left[\frac 1 4 {\text tr} 1
 \text{Tr} \ln {\cal W}_{-1}[\widehat{a}^{\dag}]
\right] {\exp}\left[{\text Tr} \ln
\left[1 - {S}^{cov}_{F} \ast \left(\gamma^{-} {A}_{-} +
\gamma^{k} {A}_{k}\right)\right]\right] .
\end{eqnarray}
So far we have used only algebraic properties, specially when
keeping track of the various combinatorical factors. Now we may
use different arguments connected with the closed loop
integrals. We observe that the term $\text{Tr} \ln {\cal
W}_{-1}[\widehat{a}^{\dag}] $ contains the integrals of
type
\begin{equation}
\int dk \  \left[\frac{1}{k}\right]_{\text CPV}
\ \left[\frac{1}{k- p_1}\right]_{\text CPV} \ldots
\left[\frac{1}{k- p_n}\right]_{\text CPV} = 0 \ \ {\text for
all} p_i \neq 0.
\end{equation}
Thus when $k_{-}$ momenta of all insertions of $A_{-}$ are
cut at some nozero value, then the non-covariant term is
field independent may be omitted. In this way we have shown
that in the LF perturbative calculation the S-matrix elements
are the same as those calculated with covariant (equal-time)
rules.

\section{Discussion and perspectives}

In this paper we have presented the canonical quantization
procedure, formulated on a single light-front, applied to the
LF-Weyl gauge QED. Special attention was paid to the free
gauge field propagators, where the ML prescription arose quite
naturally for their spurious poles. This prescription was
attributed to the presence of ghostlike modes with negative
metric. Further implementing the Gauss law weakly as a condition
on states, we have separated the physical subspace with positive
semi-definite metric. In this physical subspace the Poincar\'{e}
covariance was recovered. In the 
analysis of independent modes IR singularities were encountered
and several regularizations were studied - all of them have lead to
the same gauge field propagators in 3 + 1 dimensions. This
singularity, specific to the LF formulation, appeared for
dimensions $2 < D \leq 4$ but not for $ D = 2$ where the gauge field
sector is free from IR problems.\\
We have noticed similarity between the LF-Weyl gauge in the LF
approach and the temporal gauge $A_{0} = 0$ in the equal-time
formulation, which is very encouraging but we are also aware of
the limitations. First all this happens for the Abelian gauge
theory coupled to fermion 
currents, second there are no self-interactions with derivative
couplings for gauge fields. Unfortunately, for other types of
interaction the light-front quantization would be much more
complicated. Generally one can have nontrivial q-numbered
commutators between the vector gauge fields and the matter
fields and these commutators would depend on interactions
(contain coupling constants) so the proper perturbative rules
could be quite different from those inferred from the free
field model. Such things happen in the Abelian model with scalar
matter fields, where the LF-Weyl gauge leads to the 
equal-$x^{+}$ commutators between scalar matter fields and
vector gauge fields 
which depend on the coupling constant. Similar situation should 
also occur for the non-Abelian case, where the transverse
components $A^a_\perp$ can have q-numbered commutators with the
longitudinal ones $A^a_{-}$. Such phenomena are usually
absent in the equal-time approach and one sees that both formalisms
may be quite different for physically relevant models. \\ 
Though the above observations form a rather narrow limits for the
future LF investigations of the LF-Weyl gauge models, some possibilities
still remain. When the non-Abelian gauge fields are analyzed in
the finite volume approach (DLCQ) \cite{PauliBrodsky} the 
LF-Weyl gauge can be effectively imposed for the zero mode
gauge fields. Also it seems worth to check the another IR
regularization for the light-front procedure, where
compactification is introduced only for transverse space coordinates
$x_\perp$ - this, contrary to the common DLCQ method,
would leave the parity symmetry $x^{+} \leftrightarrow
x^{-}$ unbroken. For QED one can formally show equivalence of
perturbative covariant (equal-time) Feynman rules and the LF
rules, provided the appropriate regularization for
$(k_{-})^{-n}$ poles are introduced. Therefore the
renormalizability of QED in the LF-Weyl gauge should be checked
within the 'old-fashioned' Hamiltonian perturbation theory
\cite{Mustaki1992}. \\

\setcounter{equation}{0}

\appendix

\section{Notation}

\subsection{ Light-front coordinates in D dimensions}

In D dimensions we define 2 longitudinal coordinates
$x_L = \left(x^\pm 
= \frac{x^{0} \pm x^1}{\sqrt{2}}\right)$ and take $x^{+}$ as the
parameter of dynamical evolution. We denote transverse components
$x_\perp = (x^2, \ldots,  x^D)$ by the Latin indices $(i, j, \ldots)$.
Similarly we define components of any 4-vector. The metric has
nonvanishing components $g_{+-} = 1 , g_{ij} = - \delta_{ij}$
and the scalar product for any 4-vectors decomposes as $A \cdot
B = A^{+} B^{-} + A^{-} B^{+} - A^i B^i$.  Partial derivatives are
defined as $\partial_{\pm} = \partial / \partial x^{\pm}, \
\partial_i = \partial / \partial x^i$. Tensor components are
defined analogously e.g. $T^{\pm \mu} = \frac{1}{\sqrt{2}}
(T^{0\mu} \pm T^{1\mu})$ and summation over repeated indices is
understood. \\ 
Also we introduce the vector notation for coordinates $\vec{x} =
(x^{-}, x_\perp)$, which parameterize the light-front surface $x^{+} =
const.$ and for momenta associated with them $\vec{k} = (k_{-},
k_\perp)$. The product of such 3-vectors decomposes as
$\vec{k} \cdot \vec{x} = k_{-} x^{-} - k_i x_i$. This vector
notation is also used for the components of vector gauge field
$\vec{A} = (A_{-}, A_i)$\\

\subsection{Dirac matrices}\label{Diracmatappend}\label{Appfer}

The Dirac matrices $\gamma^\mu$ satisfy anticommutation
relation 
\begin{eqnarray}
\gamma^\mu \gamma^\nu + \gamma^\nu \gamma^\mu = 2 g^{\mu \nu}
\ ,
\end{eqnarray}
where their components are defined analogously to coordinates
e.g. $\gamma^\pm = \frac{\gamma^{0} \pm \gamma^1}{\sqrt{2}}$.
Thus $\gamma^{\pm}$ are nilpotent matrices $\left(\gamma^{\pm}
\right)^2 = 0$.\\
For the projection operators  $
\Lambda_{\pm} = {1}/{\sqrt{2}} \gamma^0 \gamma^{\pm} =
1/2 \gamma^\mp \gamma^{\pm}\ ,  $
 we have useful relations
\begin{eqnarray}
\Lambda_{\pm} \Lambda_{\pm} & = & \Lambda_{\pm}\ , \ 
\Lambda_{+} + \Lambda_{-}  =  1\ , \\
\Lambda_{+} \Lambda_{-} & = & \Lambda_{-} \Lambda_{+} =  
\gamma^{\pm} \Lambda_{\mp} =  \Lambda_{\pm} \gamma^{\mp} = 0\ ,\\
\gamma^{\pm} \Lambda_{\pm} & = & \Lambda_{\pm} \gamma^{\pm} \,\
\gamma^{0} \Lambda_{\pm} =  \Lambda_{\mp} \gamma^{0}\ ,\ \
\gamma^{i} \Lambda_{\pm} = \Lambda_{\pm} \gamma^{i}.
\end{eqnarray}
We also use the standard non-relativistic notation \cite{BjorkenDrell1964} 
$ \gamma^0  =  \beta\ ,\ \ \gamma^0 \gamma^i =  \alpha^i. $

\subsection{ Green functions}\label{AppGreen}


\noindent We define the noncovariant Feynman Green functions 
$E^1_F(x_L)$ and $E^2_F(x_L)$ as
\begin{eqnarray}
 E^1_F(x_L) & \stackrel{df}{=} & i\int_0^{\infty} \frac{dk_{-}}{2 \pi}
 \ \left[\Theta(x^{+})e^{-i {k_{-}}x^{-}} -
\Theta(-x^{+})e^{i {k_{-}}{x}^{-}}\right] = \frac{1}{2 \pi}
 \frac{1}{x^{-} -i \epsilon \ {\rm sgn}(x^{+})}\\
 E^2_F(x_L) & \stackrel{df}{=} & - x^{+} E^1_F(x_L) = -
\frac{1}{2 \pi} \frac{x^{+}}{x^{-} -i \epsilon \ {\rm sgn}(x^{+})}
\end{eqnarray}
These functions can be also represented by the 2-dimensional
Fourier integrals 
\begin{eqnarray}
E^1_F(x_L) 
& = & 
- \int_{-\infty}^{\infty} \frac{d^2{k_L}}{(2 \pi)^2} 
 \frac{e^{-i {k_L}\cdot {x_L}}}{k_{+} + i\epsilon \ {\rm sgn}\ (k_{-})}
 \label{kspace11E1}\\
 E^2_F(x_L) & = & i
\int_{-\infty}^{\infty}
\frac{d^2{k_L}}{(2 \pi)^2}  \frac{ e^{-i 
{k_L}\cdot x_L}}{[k_{+} + i \epsilon \ {\rm sgn}\ (k_{-})]^2}
\label{kspace11E2}
\end{eqnarray}
and they satisfy obvious  relations: $E^2_F(0, x^{-}) = 0, \
\partial_{+} E^2_F(x_L) = - E^1_F(x_L)$.\\ 
\noindent In $D= d+2$ dimensions we have the covariant 
massive Feynman Green function
\begin{eqnarray}
\Delta_F^{d+2}(x,m^2) & \stackrel{df}{=}&  \int_{-\infty}^{\infty}
\frac{d^dk_\perp}{(2 \pi)^d} \int_{0}^{\infty}
\frac{dk_{-}}{2\pi \ 2k_{-}} \ \left[\Theta(x^{+})e^{-i
{k}\cdot {x}} + \Theta(-x^{+})e^{+i {k}\cdot
{x}}\right]_{k_{+} = \frac{k_\perp^2 + m^2}{2k_{-}}}\nonumber\\
& = & i \int \frac{d^{d+2}{k}}{(2 \pi)^{d+2}} \frac{e^{-i {k}\cdot
x}}{k^2 - m^2 + i \epsilon}.
\end{eqnarray}
and its massless limit we denote by $D_F^{d+2}(x)
\stackrel{df}{=}\lim_{m\to 0} \Delta_F^{d+2}(x,m^2)$.
The combination of Green functions, which appears in the
expression for the longitudinal components of propagator, can be
written either as 
the finite integral over $x^{+}$ coordinate or as the Fourier
integral 
\begin{eqnarray}
\left[ 2 \partial_{-}^x D_F^{d+2}(x) + E_F^{1}(x_L)\delta^d(x_\perp)
\right] & = & -i \int_{-\infty}^{\infty}
\frac{d^dk_\perp}{(2 \pi)^d} \int_{0}^{\infty}
\frac{dk_{-}}{2\pi} \ \left[\Theta(x^{+})
\left(e^{-i\frac{k_\perp^2}{2k_{-}}x^{+}} -1\right)e^{-i
\vec{k}\cdot \vec{x}} \right. \nonumber\\ 
+ \left.\Theta(-x^{+}) \left(e^{i \frac{k_\perp^2}{2k_{-}}x^{+}}
-1\right) e^{i \vec{k}\cdot \vec{x}}\right]
& = & \Delta_\perp \int_0^{x^{+}} d\xi
D^{d+2}_F(\xi, \vec{x})\\ 
& = &  \ \Delta_\perp \int \frac{d^{d+2}{k}}{(2 \pi)^{d+2}}
\frac{e^{-i {k}\cdot x}}{k^2 + i \epsilon} \frac{1}{k_{+} + i
\epsilon' {\rm sgn}(k_{-})}.
\end{eqnarray}
Similar expressions appear for the massive case in 3+1 dimensions
\begin{eqnarray}
\left[ 2 \partial_{-}^x \Delta_F^{4}(x) + E_F^{1}(x_L)\delta^2(x_\perp)
\right] & = & 
(\Delta_\perp-m^2) \int_0^{x^{+}} d\xi
\Delta^{4}_F(\xi, \vec{x})\\ 
& = &  \ (\Delta_\perp -m^2) \int \frac{d^{4}{k}}{(2 \pi)^{4}}
\frac{e^{-i {k}\cdot x}}{k^2 - m^2 + i \epsilon} \frac{1}{k_{+} + i
\epsilon' {\rm sgn} (k_{-})}.
\end{eqnarray}
The analytically regularized inverse Laplace operator
in $d = 2$ dimensions 
\begin{equation}
[\Delta_{2 - 2 \delta}]^{-1}(x_\perp) = (-1)^{1-\delta} \int
\frac{d^2k_\perp}{(2\pi)^2} \frac{e^{i k_\perp \cdot
x_\perp}}{[k_\perp^2]^{1-\delta}} = - \frac{1}{4 \pi}
\frac{\Gamma(\delta)}{\Gamma (1-\delta)} 
\left(- \frac{4}{x^2_\perp}\right)^{\delta} 
\end{equation}
evidently is singular when $\delta \rightarrow 0$, however its
partial derivative 
\begin{equation}
\partial_i[\Delta_{2- 2\delta}]^{-1}(x_\perp) =  i (-1)^{1-\delta} \int
\frac{d^2k_\perp}{(2\pi)^2} \frac{k_i}{[k_\perp^2]^{1-\delta}} e^{i
k_\perp \cdot x_\perp} = \frac{1}{2 \pi}
\frac{\Gamma(1+\delta)}{\Gamma (1-\delta)}
\left(- \frac{4}{x^2_\perp}\right)^{\delta} \frac{x_i}{x_\perp^2} 
\end{equation}
is already finite for $\delta \rightarrow 0$
\begin{equation}
\partial_i[\Delta_{2}]^{-1}(x_\perp) =
\partial_i[\Delta_{\perp}]^{-1}(x_\perp) - i \int 
\frac{d^2k_\perp}{(2\pi)^2} \frac{k_i}{k_\perp^{2}} e^{i
k_\perp \cdot x_\perp} = \frac{1}{2 \pi} \frac{x_i}{x_\perp^2}.
\end{equation}
In Section 3 many convolutions of various Green functions
appear and they are all denoted by $*$ without reference to
their variables - hopefully leading to no misunderstandings.
Here are some explicit examples of them
\begin{eqnarray}
[ \Delta^d_\perp ]^{-1}*\Pi (\vec{x}) & = &\int d^dy_\perp
[ \Delta^d_\perp ]^{-1}(x_\perp - y_\perp)\Pi (x^{-},
y_\perp)\nonumber
\end{eqnarray}
 \begin{eqnarray}
[ \Delta^d_\perp ]^{-1}* \left\{ 2 \partial_{-}^x D_F^{d+2}(x-y) +
E_F^{1}({x}_L-{y}_L)\delta^d(x_\perp-y_\perp) \right\} = \int d^{d}w_\perp
[ \Delta^d_\perp ]^{-1}(x_\perp-w_\perp)\nonumber\\ 
\ \ \left\{ 2 \partial_{-}^x D_F^{d+2}(x_L-y_L, w_\perp -y_\perp)
+ E_F^{1}(x_L-y_L)\delta^d(w_\perp-y_\perp) \right\} \nonumber.
\end{eqnarray}

\setcounter{equation}{0}

\section{Nonlocal and massive regularizations}\label{Appreg}

Strictly in $d=2$ dimensions the integral operator
$\partial_{i} [\Delta_\perp]^{-1}(x_\perp)$ exists and the
parameterization for $A_{i}(x)$ 
\begin{equation}
A_{i}  =  C_i - {\partial_i}[\Delta_\perp]^{-1}*(\Pi +
2 \partial_j C_j)\label{modfromAitoCi}
\end{equation}
is regular, but one must be careful not to integrate the
partial derivative $\partial_{i}$ by parts because
$[\Delta_\perp]^{-1}(x_\perp)$ is ill-defined. 
For $A_{-}$ field one can introduce the analytical
regularization\footnote{See Appendix A for the definition of
$[\Delta_{2-2\delta}]^{-1}$} 
\begin{equation}
A_{-} = C_{-} - \partial_{-}[\Delta_{2-2\delta}]^{-1}*(\Pi + 
2 \partial_j C_j)\label{modfromAmtoCm}
\end{equation}
by rescaling  $A_{-}$ field
\begin{equation} 
A_{-}(x) \rightarrow [\Delta_{2\delta}]^{-1}*A_{-}(x) 
\end{equation}
with $1> \delta > 0$,  in the Lagrangian density (3.1), but
without changing the source term $A_{-}j^{-}$. Further, without
going into details of the canonical procedure, we
will present some important points which are different from the
respective
results in Section 3. The canonical 
commutator for the pair $(A_{-}, \Pi)$ is changed 
\begin{equation}
\left [\Pi (x^{+},\vec{x}), A_{-}(x^{+},\vec{y}) \right]  =  -
i \Delta_\perp [\Delta_{2-2\delta}]^{-1}(x_\perp - y_\perp)
\delta({x^-} - y^-).
\end{equation}
and the independent modes vector fields are given by 
\begin{eqnarray}
C_{i} & = & A_i - \partial_i[\Delta_{\perp}]^{-1} * \left(\Pi
+ 2 \partial_j A_j\right)\label{regfromAitoCi}\\ 
C_{-} & = & A_{-} - \partial_{-}[\Delta_{2-2\delta}]^{-1}*
\left( \Pi + 2 \partial_j A_j\right)\label{regfromAmtoCm}.
\end{eqnarray}
Also one commutator for independent modes  is changed
\begin{equation}
\left [\Pi (x^{+},\vec{x}), C_{-}(x^{+},\vec{y}) \right] = -
i \Delta_\perp [\Delta_{2-2\delta}]^{-1}(x_\perp - y_\perp)
\delta({x^-} - y^-)\label{regDBPiCm}.
\end{equation}
One may worry that these modified commutators violate causality 
because they do not vanish for spatially separated points (in
the transverse $x_\perp$ coordinates). In the free 
field case, one can easily check such observables as components of
the field strength $F_{\mu \nu}$. First one expresses them 
in terms of independent modes 
\begin{eqnarray}
F_{+-} & = & - \Delta_\perp [\Delta_{2 - 2\delta}]^{-1} *
\partial_j C_j\\
2 \partial_{-} F_{+j} & = & \left(\delta_{ij} \Delta_\perp - 2
\partial_i \partial_j \right) C_j.
\end{eqnarray}
and then finds 
 \begin{equation}
\left[2\partial_{-} F_{+i} (x^{+},\vec{x}),
F_{+-}(x^{+},\vec{y}) \right] = - i
\partial_i^x(\Delta_\perp)^2 [\Delta_{2-2\delta}]^{-1}(x_\perp
- y_\perp) \delta({x^-} - y^-)\label{FpiFpmcom}
\end{equation}
which means that also for the observables causality is violated
when $\delta \neq 0$. Therefore one has the interpretation of this
analytical regularization as a mere mathematical trick without
deeper physical meaning. Keeping this in mind, one can proceed
further with the canonical procedure 
and define the Fourier representation for $C_{-}$ and $\Pi$
free fields 
\begin{eqnarray}
C_{-}(\vec{x}) & = & \int_{-\infty}^{\infty}
\frac{d^2k_\perp}{(2 \pi)^2} \int_{0}^{\infty}
\frac{dk_{-}}{2\pi}\ \left[e^{-i \vec{k}\cdot
\vec{x}}a(\vec{k}) + e^{+i \vec{k}\cdot \vec{x}} 
a^{\dagger}(\vec{k})\right]\\ 
\Pi(\vec{x}) & = &\int_{-\infty}^{\infty} \frac{d^2k_\perp}{(2
\pi)^2} \int_{0}^{\infty} \frac{dk_{-}}{2\pi} \ \left[e^{-i
\vec{k}\cdot \vec{x}} p(\vec{k}) + e^{+i \vec{k}\cdot \vec{x}}
p^{\dagger}(\vec{k})\right]
\end{eqnarray}
which leads to the commutation relations for the creation and
annihilation operators 
\begin{equation}
\left[ a(\vec{k}), p^{\dagger}(\vec{k'})\right]  =  
\left[ a^{\dagger}(\vec{k}), p(\vec{k'})\right] =
i (2\pi)^{3} (k_\perp^2)^\delta \delta^{3}(\vec{k} -
\vec{k'})\label{comrelcp1},
\end{equation}
where again  $a^{\dag}(\vec{k})$ and $p^{\dag}(\vec{k'})$
operators may create ghostlike states with negative metric.\\
The perturbative propagators are given by the
chronological products of free fields 
\begin{eqnarray}
\langle 0 | T\ C_{-}(\vec{x})\Pi(\vec{y}) |0\rangle & = & 
E^1_F(x_L-y_L)\ \Delta_\perp
[\Delta_{2-2\delta}]^{-1}(x_\perp - y_\perp) \\
\langle 0 | T\ C_{i}({x}) C_j({y})
|0\rangle & = &  \delta_{ij} D_F^{4}(x-y).
\end{eqnarray}
and for the primary gauge fields  one obtains the general result
\begin{eqnarray}
\langle 0 | T\ A_{\mu}({x}) A_{\nu}({y})|0\rangle & = &
i \int \frac{d^{4} k}{(2 \pi)^{4}} \frac{e^{-i {k}\cdot
({x}-{y})}}{k^2 + i \epsilon} \left[ g_{\mu \nu} -
(k_\perp^2)^{\delta} \frac{\tilde{k}_\mu n_{\nu} + \tilde{k}_\nu
n_{\mu}}{k_{+}  + i\ \epsilon' \ {\rm sgn}(k_{-})} \right]
\end{eqnarray}
where the scaled momentum $\tilde{k}_\mu = (k_{+},
(k_\perp^2)^{\delta} k_{-}, k_i)$ shows that for 
$\delta > 0$ the dimensions of fields $A_{-}$ and $A_i$ are different.
Here again one can easily check 
the integral properties for propagators
\begin{eqnarray}
\langle 0 | T \ A_{-}({x}) A_{j}({0})|0\rangle 
& = & \partial_j \Delta_\perp \int_0^{x^{+}} d\xi
[\Delta_{2-2\delta}]^{-1} * D^{4}_F(\xi, \vec{x})\\ 
\langle 0 | T\ A_{-}({x}) A_{-}({0})|0\rangle  
& = & 2 \partial_{-}  (\Delta_\perp )^2
\int_0^{x^{+}} d\xi [\Delta_{2-2\delta}]^{-1}
*[\Delta_{2-2\delta}]^{-1}* D^{4}_F(\xi, \vec{x}) . 
\end{eqnarray}



The explicit mass term $ \frac 1 2 m^2 A^\mu A_\mu$ breaks gauge
invariance in the Lagrangian density which no longer describes a
bona fide gauge theory. However if this  mass is
treated only as a regularization parameter which will be finally
pushed to zero for  IR finite objects, then 
one can take the LF-Weyl gauge and start with  the Lagrangian density
\begin{equation}
{\cal L}^m_{Weyl}  =  \partial_{+} {A}_i (
\partial_{-} {A}_i - \partial_i {A}_{-}) + \frac 1 2
\left(\partial_{+} {A}_{-} \right)^2 - \frac 1 4 \left(
\partial_i {A}_j- \partial_j {A}_i\right)^2 
 - \frac{ m^2}{2} A_i^2 + {A}_{-} {j}^{-} +  {A}_i j^i. \label{mAbelLagr}
\end{equation}
The canonical momenta and commutators for
primary canonical variables are the same as in the massless theory and
the canonical Hamiltonian density has only one extra term $+
\frac{ m^2}{2} A_i^2$. Now the canonical fields 
decompose into independent modes by the regular invertible
transformations 
\begin{eqnarray}
A_{i} & = & C_i + {\partial_i}[m^2 - \Delta_\perp]^{-1}*\Pi
\label{fromAitoCim}\\ 
A_{-} & = &  C_{-} -  \partial_{-}[m^2 - \Delta_\perp]^{-1}*
\left[ 2 \partial_j C_j +  \Delta_\perp [m^2 - \Delta_\perp]^{-1}
\Pi\right] \label{fromAmtoCmm}.
\end{eqnarray}
The basic difference between the massless and the
massive cases is the equation for the longitudinal vector field 
$C_{-}$
\begin{equation}
\partial_{+} C_{-}  =  m^2 [m^2 - \Delta_\perp]^{-1}*\Pi + 
[m^2 - \Delta_\perp]^{-1}* \left[\partial_i j^i 
-  \Delta_\perp [m^2 - \Delta_\perp]^{-1}* \partial_{-} j^{-}\right].
\end{equation} 
which, for free fields, leads to the Fourier
representation with the explicit $x^{+}$ dependence 
\begin{eqnarray}
{\Pi}(\vec{x}) & = &\int_{-\infty}^{\infty} \frac{d^2k_\perp}{(2
\pi)^2} \int_{0}^{\infty} \frac{dk_{-}}{2\pi}
\ \left[e^{-i \vec{k}\cdot
\vec{x}} p(\vec{k}) + e^{+i \vec{k}\cdot \vec{x}}
p^{\dagger}(\vec{k})\right]\\
{C_{-}}(x^{+},\vec{x}) & = & - x^{+} \ m^2 [m^2 -
\Delta_\perp]^{-1}*\Pi (\vec{x}) \nonumber\\
&+ &\int_{-\infty}^{\infty} \frac{d^2k_\perp}{(2
\pi)^2} \int_{0}^{\infty} \frac{dk_{-}}{2\pi}
 \ \left[e^{-i \vec{k}\cdot
\vec{x}} c_{-}(\vec{k}) + e^{+i \vec{k}\cdot \vec{x}}
c^{\dagger}_{-}(\vec{k})\right]
\end{eqnarray}
and one gets nonzero  longitudinal components of the propagator
\begin{equation}
<0| T\ C_{-}(x) C_{-}(y) | 0>  =  m^2 E_F^2(x_L - y_L)
[m^2 - \Delta_\perp]^{-1}(x_\perp-y_\perp).
\end{equation}
Also the propagator for $A_{-}$ fields is modified
\begin{eqnarray}
\hspace{-10pt}&& <0| T\ A_{-}(x) A_{-}(y) | 0>  =   m^2
E_F^2(x_L - y_L) [m^2 -
\Delta_\perp]^{-1}(x_\perp-y_\perp)\nonumber\\ 
&&\ \ \ +  2 \partial^x_{-}
\Delta_\perp [m^2 - \Delta_\perp]^{-2} *\left[ 2 \partial_{-}^x
\Delta_F^{2+2}(x-y)  + E_F^1(x_L - y_L) \delta^2(x_\perp -
y_\perp)\right]  \nonumber\\
&  &\ \ \ =  2 \partial_{-}^x \int_0^{x^+ -y^{-}} d\xi \Delta_F^{2+2}(\xi,
\vec{x} - \vec{y})  +  m^2 \int_0^{x^+ -y^{-}} d\xi
\int_{0}^{\xi} d\eta \Delta_F^{2+2}(\eta, \vec{x} - \vec{y}).
\end{eqnarray}
Finally one derives the Fourier representation  of gauge field propagator
\begin{eqnarray}
<0| T\ A_{\mu}(x) A_{\nu}(0) | 0> & =& i \int
\frac{d^4k}{(2\pi)^4} \frac{e^{i k\cdot x}}{k^2 - m^2 + i
\epsilon} \left( - g_{\mu \nu} + \frac{k_\nu n_\mu + 
k_\mu n_\nu}{k_{+} + i \epsilon'{\rm sgn}(k_{-}) } \right.\nonumber\\
&& + m^2 
\left. \frac{n_\mu n_\nu}{[k_{+} + i \epsilon'{\rm sgn}(k_{-})]^2}
\right)
\end{eqnarray}
where the $n_\mu n_\nu$ term is characteristic for the
massive regularization and it smoothly vanishes in the limit
$m^2 \rightarrow 0$.

\setcounter{equation}{0}

\section{Poincar\'{e} generators}

The analysis of the Poincar\'{e} covariance starts with the definition
of the canonical energy-momentum tensor 
\begin{equation}
{\cal T}^{\mu \nu} = \partial^\nu A_i \frac{\partial\ {\cal
L}}{\partial \ \partial_\mu A_i} + \partial^\nu A_-
\frac{\partial\ {\cal L}}{\partial \ \partial_\mu A_-} - g^{\mu
\nu} {\cal L} = - F^{\mu \lambda} \partial^\nu A_\lambda -
g^{\mu \nu} {\cal L}\label{defcanten} 
\end{equation}
where we take the Lagrangian density (\ref{AbelLagr}) with  the
LF-Weyl gauge condition explicitly implemented  and we confine our 
discussion to the free Abelian theory in 3+1 dimensions. This
canonical tensor may be expressed  in terms of the symmetric
energy-momentum tensor 
\begin{equation}
\Theta^{\mu \nu} = F^{\mu \lambda} F_{\lambda}^{\ \nu} - g^{\mu
\nu} {\cal L} \label{defsymten}
\end{equation}
and one gets
\begin{equation}
{\cal T}^{\mu \nu} = \Theta^{\mu \nu} - \partial_\lambda (F^{\mu
\lambda} A^{\nu}) + \delta^{\mu +} A^\nu G \label{cantosymT}
\end{equation}
where $G$ is the Gauss law operator in 3+1 dimensions. Similarly
we define the canonical angular momentum tensor
\begin{equation}
{\cal M}^{\mu \nu \rho} = {\cal T}^{\mu \rho} x^{\nu} - {\cal
T}^{\mu \nu} x^{\rho} + \frac{\partial\ {\cal
L}}{\partial \ \partial_\mu A_\nu} A^\rho  - \frac{\partial\ {\cal
L}}{\partial \ \partial_\mu A_\rho} A^\nu
\end{equation}
and then the generators of the Poincar\'{e} transformations 
\bml
\begin{eqnarray}
P^\mu & = & \int d^2x_\perp dx^{-} \ {\cal T}^{+\mu}\\
M^{\mu \nu} & = & \int d^2x_\perp dx^{-} \ {\cal M}^{+\mu \nu}.
\end{eqnarray}
\eml
We notice that the relation (\ref{cantosymT}) allows to write
the expressions for generators as
\begin{eqnarray}
P^\mu & = & \int d^3\vec{x} \ {\cal T}^{+\mu} = \int
d^3\vec{x} \ \left[ \Theta^{+\mu}+ A^\mu G\right]\\ 
M^{\mu \nu} & = & \int d^3\vec{x} \ {\cal M}^{+ \mu \nu} = 
\int d^3\vec{x} \ \left[ ({\Theta}^{+ \mu} + A^\mu G) x^{\nu} -
({\Theta}^{+\nu} + A^{\nu} G)x^{\mu}\right].
\end{eqnarray}
where the expected form has regular modifications by the
noncovariant terms $A^\mu G$. These terms arise from the 
noncovariant gauge condition and they generate $x^{+}$
dependence of the Poincar\'{e} generators $M^{+i}$
\begin{equation}
\frac{d }{dx^{+}} M^{+i} =  \int d^2x_\perp dx^{-} \
G A^{i} \label{nonindepMpi}
\end{equation}
while other generators are $x^{+}$ independent. For the
quantum theory one needs some prescription for noncommuting
operators and we choose symmetric product for $A^\mu G$ and
normal ordering for operators appearing in $\Theta^{\mu \nu}$. \\ 
After a tedious but straightforward calculation we find all
commutators that we need. First for the Poincar\'{e} generators and
the independent vector fields $ \vec{A}_\mu =
(A_{-},A_i)$ \footnote{The momentum field $\Pi$
transforms according to the relation $\Pi = \partial_+ A_{-} -
\partial_i A_i$ and therefore it will be discarded hereafter.}
one gets
\begin{eqnarray}
\left[ P^\mu, \vec{A}^\mu(x) \right] & = & - i \partial^\mu
\vec{A}^\mu(x)\\ 
\left[M^{\mu \nu}, \vec{A}^\lambda(x)\right] & = & -
i\left(x^{\mu}\partial^\nu - x^\nu 
\partial^{\mu}\right) \vec{A}^\lambda(x) - i\left(g^{\mu \lambda}
\vec{A}^\nu - g^{\nu \lambda} \vec{A}^\mu\right)(x)
\end{eqnarray}
so $\vec{A}^\mu$ transforms covariantly. The noncovariant
behaviour appears for the field strength 
tensor $F^{\mu \nu}$, where beside the desired relations 
\begin{eqnarray}
\left[ P^\mu, F^{\lambda \rho}(x)\right] & = & - i \partial^\mu
F^{\lambda \rho}(x)\\ 
\left[M^{\mu \nu}, F^{\lambda \rho}(x)\right] & = & -
i\left(x^{\mu}\partial^\nu - x^\nu \partial^{\mu}\right) F^{\lambda
\rho}(x)\nonumber\\ 
& + & i g^{\nu \lambda} F^{\mu \rho}(x) + i g^{\nu \rho}
F^{\lambda \mu}(x) - i g^{\mu \lambda} F^{\nu \rho}(x) 
- i g^{\mu \rho} F^{\lambda \nu}(x)
\end{eqnarray}
one also gets the anomalous commutator\footnote{Here
$[\partial_{-}]^{-1}$ may be any real valued Green function.} 
\begin{equation}
\left[M^{+i}, F^{-j}(x)\right]  =  - i\left(x^{+}\partial^i - x^i
\partial^{+}\right)F^{-j}(x) - i F^{ij}(x)
 -  i g^{ij}\left( F^{+-}(x) + \frac 1 2
[\partial_{-}]^{-1}*G(x)\right)
\end{equation}
which is a direct consequence of (\ref{nonindepMpi}). Fortunately the
components  $F^{- j}$ are not present in the Poincar\'{e}
generators and  one still has covariant  commutators
\begin{eqnarray}
\left[ M^{\mu \nu}, \Theta^{+\lambda}(x)\right] & = & -
i\left(x^{\mu}\partial^\nu - x^\nu \partial^{\mu}\right)
\Theta^{+\lambda}(x)\nonumber\\ 
& + & i g^{\nu +} \Theta^{\mu \lambda}(x) + i g^{\nu \lambda}
\Theta^{+ \mu }(x) - i  g^{\mu +}\Theta^{\nu \lambda}(x) 
- i  g^{\mu \lambda}\Theta^{+ \nu}(x). 
\end{eqnarray}
Finally for the $ (G \vec{A}^\rho + \vec{A}^\rho G)$
term one finds
\begin{eqnarray}
\left[ M^{\mu \nu},  (G \vec{A}^\rho + \vec{A}^\rho
G)(x)\right] & = & - i\left(x^{\mu } \partial^{\nu} - x^{\nu}
\partial^{\mu}\right) (G \vec{A}^\rho + \vec{A}^\rho G)(x)
\nonumber\\  
& - & i\left(g^{\mu +} g^{\nu -} - g^{\nu +} g^{\mu -}\right)
(G \vec{A}^\rho + \vec{A}^\rho G)(x) 
\nonumber\\ 
& - & i \left[g^{\mu \rho} (G \vec{A}^\nu + \vec{A}^\nu G)(x) -
g^{\nu \rho}  (G \vec{A}^\mu + \vec{A}^\mu G)(x)\right] .
\end{eqnarray}
Gathering all these results one obtains the commutator algebra
for Poincar\'{e} generators (\ref{PoinccomMP}) and
(\ref{PoinccomMM}).

\end{document}